\documentstyle[12pt]{article}
\begin{document}

\tolerance=5000
\def\pp{{\, \mid \hskip -1.5mm =}}
\def\cL{{\cal L}}
\def\be{\begin{equation}}
\def\ee{\end{equation}}
\def\bea{\begin{eqnarray}}
\def\eea{\end{eqnarray}}
\def\tr{{\rm tr}\, }
\def\nn{\nonumber \\}
\def\e{{\rm e}}
\def\D{{D \hskip -3mm /\,}}

\def\SEH{S_{\rm EH}}
\def\SGH{S_{\rm GH}}
\def\AdS5{{{\rm AdS}_5}}
\def\S4{{{\rm S}_4}}
\def\gfv{{g_{(5)}}}
\def\gfr{{g_{(4)}}}
\def\SC{{S_{\rm C}}}
\def\RH{{R_{\rm H}}}

\def\wlBox{\mbox{
\raisebox{0.1cm}{$\widetilde{\mbox{\raisebox{-0.1cm}\fbox{\ }}}$}}}
\def\htBox{\mbox{
\raisebox{0.1cm}{$\hat{\mbox{\raisebox{-0.1cm}{$\Box$}}}$}}}

\ 

\vskip -3cm

\  \hfill
\begin{minipage}{2.5cm}
NDA-FP-?? \\
September 2001 \\
\end{minipage}

\vfill

\begin{center}
{\large\bf 
Anti-de Sitter Black Hole Thermodynamics in Higher Derivative Gravity 
and New Confining-Deconfining Phases in dual CFT.}

\vfill

{\sc Shin'ichi NOJIRI}\footnote{nojiri@cc.nda.ac.jp} 
and {\sc Sergei D. ODINTSOV}$^{\spadesuit}$\footnote{
On leave from Tomsk State Pedagogical University, 
634041 Tomsk, RUSSIA. \\
odintsov@ifug5.ugto.mx, odintsov@mail.tomsknet.ru} \\

\vfill

{\sl Department of Applied Physics \\
National Defence Academy,
Hashirimizu Yokosuka 239-8686, JAPAN}

\vfill

{\sl $\spadesuit$
Instituto de Fisica de la Universidad de Guanajuato, \\
Lomas del Bosque 103, Apdo. Postal E-143, 
37150 Leon,Gto., MEXICO}

\vfill

{\bf ABSTRACT}

\end{center}

The thermodynamics of d5 AdS BHs with positive, negative or zero 
curvature spatial section in higher derivative (HD) gravity is 
described.
HD contribution to free energy may change its sign which leads to more 
complicated regime for Hawking-Page phase transitions. Some variant of
d5 HD gravity is dual to ${\cal N}=2$ $Sp(N)$ SCFT up to the next-to-leading 
order
in large $N$. Then, according to Witten interpretation the stable AdS BH 
phase corresponds to deconfinement while global AdS phase corresponds to 
confinement. Unlike to Einstein gravity in HD theory the critical $N$ 
appears. It may influence the phase transition structure. In particulary,
what was confining phase above the critical value becomes the deconfining 
phase below it and vice-versa.

\newpage

\section{Introduction\label{sec1}}

It has been observed quite long ago by Hawking and Page \cite{HP}
that Anti-de Sitter (AdS) Black Holes (BHs) thermodynamics admits  
phase transitions. These Hawking-Page phase transitions occur as the 
following: low temperature BHs are not stable and global AdS 
spacetime is then preferrable state. On the same time, high 
temperature BHs are stable and they do not decay to the global AdS spacetime. 

The invention of AdS/CFT correspondence \cite{AdS} increased 
the interest to phase transitions for AdS BHs. Indeed, 
Witten \cite{witten} demonstrated that Hawking-Page phase 
transition corresponds to a deconfinement-confinement transition 
in the large-$N$ limit of an ${\cal N}=4$ $SU(N)$ super 
Yang-Mills theory living on the boundary of 5d AdS BH. 
This is easily seen when calculating the expectation value 
of the temporal Wilson loop operator which is an order 
parameter for the spontaneous symmetry breaking for a subgroup 
of the gauge group center. When this expectation value is not 
zero (as is the case for stable AdS BHs phase) then 
deconfinement is realized. On the contrary, when it is zero 
(global AdS space) the confinement is realized. The same 
interpretation of phase transitions \cite{b,CM,SSW} survives 
for 5d AdS BHs with zero or negative curvature for spatial 
section. 
 
In the present work we discuss (spherical, flat or hyperbolic) AdS BH 
thermodynamics and especially the role
 played by HD terms to Hawking-Page phase transitions. The free energy 
for such BHs is found and its dependence from HD terms coefficients is 
clarified. As a result, there appear several regimes for phase transitions.
Using dual $Sp(N)$ SCFT interpretation of specific HD gravity model one comes
to complicated structure of Witten confinement-deconfinement phase 
transitions. In particular, there occurence of critical $N$ is remarkable.

\section{Thermodynamics of bulk AdS black hole\label{sec2}}

In this section,  we review thermodynamics of AdS BH in bulk 
$R^2$-gravity, based on \cite{NOOlng}.  We consider 
the case that HD terms contain the Riemann 
tensor square term, i.e. $R_{\mu\nu\xi\sigma}R^{\mu\nu\xi\sigma}$. 
As it will be shown the presence of this term will lead
to interesting consequences for AdS BH thermodynamics.

The general action of $d+1$ dimensional $R^2$-gravity is given 
by 
\be
\label{vi}
S=\int d^{d+1} x \sqrt{-\hat G}\left\{a \hat R^2 
+ b \hat R_{\mu\nu} \hat R^{\mu\nu}
+ c  \hat R_{\mu\nu\xi\sigma} \hat R^{\mu\nu\xi\sigma}
+ {1 \over \kappa^2} \hat R - \Lambda 
+ L_{\rm matter}\right\}\ .
\ee
When $c=0$, Schwarzschild-anti de Sitter space is an exact 
solution of gravitational theory:
\bea
\label{SAdS}
&& ds^2=\hat G_{\mu\nu}dx^\mu dx^\nu 
=-\e^{2\rho_0}dt^2 + \e^{-2\rho_0}dr^2 
+ r^2\sum_{i,j}^{d-1} g_{ij}dx^i dx^j\ ,\nn
&& \e^{2\rho_0}={1 \over r^{d-2}}\left(-\mu + {kr^{d-2} \over d-2} 
+ {r^d \over l^2}\right)\ .
\eea
Here $l$ is the radius of the asymptotic AdS space, given by 
solving the equation
\be
\label{ll}
0={d^2(d+1)(d-3) a \over l^4} + {d^2(d-3) b \over l^4} 
- {d(d-1) \over \kappa^2 l^2}-\Lambda\ .
\ee
For non-vanishing $c$, such an S-AdS BH solution may be
constructed perturbatively. 
In this section, we only consider the case $a=b=0$ for simplicity: 
\be
\label{rie}
S=\int d^{d+1} x \sqrt{-\hat G}\left\{
c \hat R_{\mu\nu\xi\sigma}\hat R^{\mu\nu\xi\sigma}
+ {1 \over \kappa^2} \hat R - \Lambda \right\}\ .
\ee
When we assume the metric (\ref{SAdS}) with $\mu=0$, the scalar, 
Ricci and Riemann curvatures are given by
\be
\label{rr}
\hat{R}=- {d(d+1)\over l^2},\; \hat{R}_{\mu\nu}=
-{d\over l^{2}}G_{\mu\nu},\; \hat{R}_{\mu\nu\xi\sigma} 
= -{1\over l^2}\left( \hat{G}_{\mu\xi}\hat{G}
_{\nu\sigma} -\hat{G}_{\mu\sigma}\hat{G}_{\nu\xi} \right), 
\ee
which tell that the curvatures are covariantly constant. The 
equation of the motion derived from the action (\ref{rie}) (no matter)
is:
\be
\label{cl}
0= -{\hat{G}_{\zeta\xi} \over 2}\left\{ 
c \hat{R}_{\mu\nu\rho\sigma}
\hat{R}^{\mu\nu\rho\sigma} +{\hat{R} \over \kappa^{2}}
 -\Lambda \right\} 
+ 2c \hat{R}_{\zeta\mu\nu\rho}\hat{R}_{\xi}^{\ \mu\nu\rho}
+{\hat{R}_{\zeta \xi} \over \kappa^{2}}
+4c D_{\rho}D_{\kappa}\hat{R}_{\zeta\; \xi}^{\; \rho \; \kappa}.
\ee
Then substituting Eqs.(\ref{rr}) into (\ref{cl}), one finds the
relation between $c$, $\Lambda$ and $l$
\bea
\label{clc}
0={2c \over l^4}d(d-3)-{d(d-1) \over \kappa^{2} l^{2}}-\Lambda\ ,
\eea 
which defines the radius $l$ of the asymptotic AdS space 
even if $\mu\neq 0$. 
For $d+1=5$ with $\mu\neq 0$, using Eq.(\ref{cl}),  
we get the perturbative solution from (\ref{SAdS}), 
which looks like:
\bea
\label{dAdS}
\e^{2\rho}={1\over r^2}\left\{ -\mu +{k\over 2}r^2 + {r^{4}\over l^2}
+{2\mu^2 \epsilon \over r^{4} }\right\}, \quad 
\epsilon = c\kappa^{2}\ .
\eea
Suppose that $g_{ij}$ (\ref{SAdS}) corresponds to the Einstein 
manifold, 
defined by $r_{ij}=kg_{ij}$, where $r_{ij}$ is Ricci tensor 
defined by $g_{ij}$ and $k$ is the constant. 
For example, if $k>0$ the boundary can be three dimensional 
sphere, if $k<0$, hyperboloid, or if $k=0$, flat space. 
Properly normalizing the coordinates, one can choose $k=2$, $0$, 
or $-2$. 

After Wick-rotating the time variable by $t \to i\tau$, the
free energy $F$ can be obtained from the action $S$ (\ref{rie}) :
$F=-TS$, where the classical solution is substituted. 
Multiplying $\hat{G}^{\zeta\xi}$ to (\ref{cl}) in case 
that $D_{\rho}D_{\kappa}
\hat{R}_{\zeta\; \xi}^{\; \rho \; \kappa}={\cal O}(\epsilon)$ 
as in the solution (\ref{dAdS}), one finds for $d=4$
${1\over \kappa^{2}}\hat{R}=- {c \over 3} \hat{R}_{\mu\nu\rho\sigma}
\hat{R}^{\mu\nu\rho\sigma} + {5 \over 3}\Lambda + 
{\cal O}\left(\epsilon^2\right)$. 

Then the action (\ref{rie}) can be rewritten as
\be
\label{ri2}
S=\int d^5 x \sqrt{-\hat G}\left\{
{2 \over 3}c \hat R_{\mu\nu\xi\sigma}\hat R^{\mu\nu\xi\sigma}
+ {2 \over 3} \Lambda \right\}\ .
\ee
Since $\hat R_{\mu\nu\xi\sigma}\hat R^{\mu\nu\xi\sigma}
={40 \over l^2}+{72 \mu^2 \over r^8} + {\cal O}(\epsilon)$,
by using (\ref{clc}) with $d=4$, we obtain
\bea
\label{sr}
S &=& - \int d^{5}x \sqrt{-\hat G} \left( {8 \over \kappa^2 l^{2}}
 - {32 c \over l^4} - {48c \mu^2 \over r^8}\right) \nn 
&=& -{V_{3} \over T}\int ^{\infty}_{r_{H}} dr r^{3}
\left( {8 \over \kappa^2 l^{2} }- {32 c \over l^4} 
 - {48c \mu^2 \over r^8}\right)\ .
\eea
Here $V_{3}$ is the volume of unit 3d sphere for $k=2$ 
and we assume $\tau$ 
has a period  ${1\over T}$.  The expression for $S$ contains 
the divergence coming from large $r$. In order to subtract the 
divergence, we regularize $S$ (\ref{sr}) by cutting off the 
integral at a large radius $r_{\rm max}$ and subtracting the 
solution with $\mu =0$ in a same way as in \cite{NS}: 
\bea
\label{regS}
S_{\rm reg}&=& 
-{V_{3} \over T}\left\{ \int ^{r_{\rm max}}_{r_{H}} dr r^{3}
\left( {8 \over \kappa^2 l^{2} }- {32 c \over l^4} 
 - {48c \mu^2 \over r^8}\right) \right. \nn
&& \left. -\e^{\rho(r=r_{\rm max})-\rho(r=r_{\rm max};\mu =0) }
\int ^{r_{\rm max}}_{0} dr r^{3} 
\left( {8 \over \kappa^2 l^{2} }- {32 c \over l^4} \right)
\right\}\ .
\eea
The factor $\e^{\rho(r=r_{\rm max})-\rho(r=r_{\rm max};\mu =0)}$ 
is chosen so that the proper length of the circle which 
corresponds to the period ${1 \over T}$ in the Euclidean 
time at $r=r_{\rm max}$ coincides with each other in the 
two solutions. Taking $r_{\rm max} \to \infty$, one finds 
\bea
\label{free1}
F= V_{3}\left\{\left( {l^2 \mu \over 8} 
 - {r_H^4  \over 4}\right)
\left( {8 \over \kappa^2 l^{2} }- {32 c \over l^4} \right)
 - {12 c\mu^2 \over r_H^4}\right\}\ .
\eea
The horizon radius $r_{H}$ is given by solving 
the equation $\e^{2\rho_0(r_H)}=0$ in (\ref{dAdS}). 
We can solve $r_{H}$ perturbatively up to first order 
on $c$ by putting $r_{H}=r_{0}+ c\delta r$, 
where $r_{0}$ is the horizon radius when $c=0$: 
\be
\label{hrznrds}
r_{H}=r_{0}-{c\mu^{2}\kappa^{2} \over r_{0}^{3}
\left( 2\mu -{k \over 2}r_{0}^{2} \right) }\ ,
\quad r_{0}^{2}=-{k l^{2} \over 4} + {1\over 2}
\sqrt{{k^2 \over 4}l^{4}+ 4\mu l^{2} }\ .
\ee
We can also rewrite the black hole mass $\mu$ (using $r_{H}$)
up to first order on $\epsilon$ $(\epsilon = c\kappa^{2})$:
$\mu = {k\over 2}r_{H}^{2}+{r_{H}^{4} \over l^{2}} 
+{2 \epsilon \over r_{H}^{4}}\left( {k \over 2}r_{H}^{2}
+{r_{H}^{4} \over l^{2} } \right)^{2}$. Then  $F$ looks like
\bea
\label{free2}
F= {V_{3} \over \kappa^{2} l^{2}} \left[ {l^{2} k \over 2}
r_{H}^{2} -r_{H}^{4}+\epsilon \left\{ {l^{2}k^{2} \over 2}
+{6r_{H}^{4}\over l^{2} } - {12 l^2 \over r_H^4}
\left( {k \over 2}r_{H}^{2}
+{r_{H}^{4} \over l^{2} } \right)^{2} \right\}\right]\ .
\eea
The Hawking temperature $T_H$ is given by 
\be
\label{ht1}
T_H = {(\e^{2\rho})'|_{r=r_{H}} \over 4\pi} 
= {1 \over 4\pi}\left\{ {4r_{H} \over l^2 }+ {k\over r_{H}}
-{8\epsilon \over r_{H}^{7} } \left({k\over 2}r_{H}^{2}
+ {r_{H}^{4} \over l^2} 
\right)^{2} \right\} \  ,
\ee
where $'$ denotes the derivative with respect to $r$.
Then the entropy ${\cal S}= -{dF \over dT_H}
=-{dF \over dr_{H}}{dr_{H} \over dT_H}$ and 
the energy $E=F+T_H{\cal S}$ have the following form:  
\bea
\label{entropy}
{\cal S }&=& {4\pi V_{3} r_{H}^{3} \over \kappa^{2} }
\left\{ 1- {1 \epsilon \over l^{2}} 
\left( -8 - {4k l^{2} \over r_{H}^{2} } 
+ {3 k^{2} l^{4} \over 2 r_{H}^{4} } \right) 
\left( 1-{k l^{2} \over 4 r_{H}^{2} } \right)^{-1} 
\right\} \ ,\\
\label{energy}
E &=& {3 V_{3} \over \kappa^{2} }
\left\{ {1\over 2}kr_{H}^{2}+{r_{H}^{4} \over l^{2}} \right. \nn
&& \left. + \epsilon \left( {34 r_{H}^{4} \over 3l^{4} }
 - {17 k r_{H}^{2} \over 6 l^{2}}
 - {19 \over 6}k^{2}+{k^{3} l^{2} \over 24 r_{H}^{2}} \right)
\left( 1-{k l^{2} \over 4 r_{H}^{2} } \right)^{-1}
\right\} 
\ .
\eea
It is remarkable that the entropy ${\cal S}$ 
is not proportional to the area of the horizon when $k\neq 0$ 
and the energy $E$ is not to $\mu$, either. We should note that 
the entropy ${\cal S}$ was 
proportional to the area and the energy $E$ to $\mu$ even in 
$R^2$-gravity if there is no the squared Riemann tensor 
term ($c=0$ in (\ref{vi})) \cite{NOOlng}, where we have 
the following expressions:
\bea
&& F= -{V_{3} \over 8}r_{H}^{2} \left( {r_{H}^{2} \over l^{2}}
 - {k \over 2} \right)\left( {8 \over \kappa^2} 
 - {320 a \over l^2} -{64 b \over l^2} \right) \; , \\
\label{ent}
&& {\cal S }={V_{3}\pi r_H^3 \over 2}
\left( {8 \over \kappa^2}- {320 a \over l^2}
 -{64 b \over l^2} \right)\ ,\quad 
E= {3V_{3}\mu \over 8}
\left( {8 \over \kappa^2}- {320 a \over l^2}
 -{64 b \over l^2} \right)\ .\nonumber
\eea
Here $a$ and $b$ are given in (\ref{vi}). Thus, we demonstrated the role 
of non-zero $c$ contribution in the thermodynamics of AdS BH. 

\section{Phase transitions in higher derivative AdS 
gravity\label{sec3}}

It has been suggested by Hawking and Page \cite{HP}, there 
is a phase transition between AdS BH spacetime 
and global AdS vacuum. BH is stable at high 
temperature but it becomes unstable at low temperature. 
 From the point of view of the AdS/CFT correspondence \cite{AdS}, 
this phase transition could correspond to the 
confinement-deconfinement transition in dual gauge theory 
\cite{witten}. In this section, we investigate the phase 
structure in HD 
gravity by using the thermodynamical quantities obtained in the 
previous section. 

Note that there is a minimum $T_{\rm min}={\sqrt{k} \over \pi l}$ 
for the Hawking temperature $T_H$ in (\ref{ht1}) when $k>0$ and 
$\epsilon=0$ or $c=0$ when the horizon radius $r_H$ is given by 
$r_H^2=r_1^2\equiv {kl^2 \over 4}$.
The existence of the minimum tells that BH cannot 
exist at low temperature $T<T_{\rm min}$. When we 
Wick-rotate the AdS metric, we can freely impose the periodic 
boundary condition for the Euclidean time variable. Then the 
temperature of the AdS can be arbitrary. Only global AdS can exist 
when $T<T_{\rm min}$. The free energy $F$ of the AdS vanishes by 
the definition here. On the other hand, the free energy of BH 
is given in (\ref{free1}), which vanishes at 
$r_H^2=r_2^2\equiv {l^2 k \over 2}$when $\epsilon=0$ or $c=0$. 
The corresponding critical Hawking 
temperature is given by $T_H=T_c={3 \sqrt{2k} \over 4\pi l}$. 
Then when $c=0$, there is a phase transition between  
AdS BH and AdS when $T_H=T_c$. When $T_H>T_c$, 
the BH free energy $F$ with $\epsilon=0$ is negative, 
and BH will be preferable. Since the phase 
transition is of the first order, BH can 
exist when $T_c>T_H\geq T_{\rm min}$ but it becomes 
unstable and it decays into global AdS. This  also proves that 
the horizon radius has a minimum when $r_H=r_2=l\sqrt{k \over 2}$. 

What happens when $\epsilon\neq 0$ but small? Then
\bea
\label{cmin}
& T_{\rm min}= {\sqrt{k} \over 4\pi}\left[{4 \over l} 
- {36\epsilon \over l^3} + {\cal O}\left(\epsilon^2\right) 
\right]\ , \quad 
& r_1^2={kl^2 \over 4} - {15k\epsilon \over 2}
 + {\cal O}\left(\epsilon^2\right) \ , \\
\label{Tcc}
& T_c= {\sqrt{2k} \over 4\pi}\left[ {3 \over l} 
+ {12 \epsilon \over l^3} + {\cal O}\left(\epsilon^2\right) 
\right]\ , \quad
& r_2^2= {l^2 k \over 2} + 28k\epsilon  
+ {\cal O}\left(\epsilon^2\right) \ .
\eea
If $\epsilon$ is positive (negative), the correction makes 
$T_{\rm min}$ smaller (larger) and makes $T_c$ larger (smaller). 

When $T_H\sim T_{\rm min}$ or $T_c$, the essential behaviors are not 
changed by the corrections. The region corresponds to $r_H\sim l$.  
We should, however, note that the behaviors when $r_H$ is small 
might be changed. In Eq.(\ref{ht1}), the Hawking 
temperature $T_H$ may become negative if $\epsilon$ is positive. 
Although $T$ should be positive, this means that the temperature 
of BH can be small for small $r_H$.  
Assuming $r_H^2={\cal O}(\epsilon)$ and $\epsilon$ is small, 
one finds $T_H$ (\ref{ht1}) vanishes when 
$r_H^2=2k\epsilon + {\cal O}\left(\epsilon^2\right)$.
Then, the free energy  (\ref{free1}) becomes 
\be
\label{srHF}
F=-{3V_3k^2 \epsilon \over 2\kappa^2 }\ ,
\ee
which is negative when $\epsilon$ is positive. 
Since $r_H^2=2k\epsilon + {\cal O}\left(\epsilon^2\right)$, 
$\epsilon$ should be positive if $k$ is positive, as we assume here. 
Then the AdS black hole spacetime might be preferred than 
AdS spacetime for small $r_H$ again. 
When $r_H^2={\cal O}(\epsilon)$, the correction becomes large and 
 the perturbation with respect to $\epsilon$ is not 
valid. The above results suggest that the essential 
behavior might be changed for small $r_H$. 

So far in this section, we have assumed $k$ should be positive. 
When $c=0$, the free energy (\ref{free2}) is always negative if 
$k$ is not positive ($k\leq 0$) then the AdS black hole is always 
stable. Furthermore, when $k<0$, the horizon radius 
(\ref{hrznrds}) does not 
vanish even if $\mu=0$:
\be
\label{rminknega}
r_H^2= r_{{\rm min}-} \equiv {|k| l^2 \over 2}
= -{k l^2 \over 2}\ .
\ee
Eq.(\ref{rminknega}) does not change even if $c\neq 0$. 
When $r_H^2= r_{{\rm min}-}^2$, the free energy (\ref{free2}) 
has the following form:
\be
\label{Fmin}
F=F_L\equiv -{V_3 l^4 k^2 \over 2\kappa^2 l^2 } \left(1 - 
{4\epsilon \over l^2}\right)\ .
\ee
Then if 
\be
\label{phph}
1 - {4\epsilon \over l^2}< 0 \ ,
\ee
the free energy can be positive. When $r_H^2= r_{{\rm min}-}^2$, 
the Hawking temperature (\ref{ht1}) has a minimum:
$T_H={\sqrt{-2k} \over 4\pi l}$.
For large $r_H$, the Hawking temperature (\ref{ht1}) and 
the free energy (\ref{free2}) are:
\be
\label{THL}
T_H={r_H \over \pi l^2}\left(1 - {2\epsilon \over l^2}\right) 
\ ,\quad F\rightarrow F_H\equiv -{V_3 \over \kappa^2 l^2}
\left(1 - {18\epsilon \over l^2}\right)\ .
\ee
Eq.(\ref{THL}) tells that if ${\epsilon \over l^2}<{1 \over 2}$, 
the large (large $r_H$) black hole is at high temperature, as is usual 
for AdS BH (of course, it is not correct for the black 
hole in the flat background). Eqs.(\ref{Fmin}), (\ref{THL}) 
indicate that there are several critical points:
\begin{enumerate}
\item If ${\epsilon \over l^2}<{1 \over 18}$, both of 
$F_L$ and $F_H$ are negative, then the black hole is 
always stable, as in $\epsilon=0$ case. For $Sp(N)$ SCFT (see next 
section) it gives $N$ more than $288$ and deconfining phase. 
 \item If ${1 \over 4}>{\epsilon \over l^2}>{1 \over 18}$, we 
find $F_L<0$ but $F_H>0$. Then at low temperature BH 
 is stable but it becomes unstable at high temperature. For dual SCFT
 one gets deconfinement-confinement transition ($64<N<288$). 
\item If ${1 \over 2}>{\epsilon \over l^2}>{1 \over 4}$, both of 
$F_L$ and $F_R$ are positive. Therefore BH becomes 
always unstable and  decays into the AdS vacuum. For dual $Sp(N)$ SCFT 
it corresponds to confining phase ($32<N<64$). 
\item When ${\epsilon \over l^2}>{1 \over 2}$, the structure 
becomes very complicated.  It could be the perturbation with 
respect to $\epsilon$ is not valid and we cannot come 
to any definite conclusion.
\end{enumerate}
Anyway the above observed phase transition does not occur 
when $c=0$ ($\epsilon=0$) but it appears when 
${\epsilon \over l^2}> {1 \over 18}$. Note, 
Eq.(\ref{THL}) does not depend on $k$, then 
if ${\epsilon \over l^2}> {1 \over 18}$, the black hole becomes 
always unstable. We should also note that the case 
${1 \over 4}>{\epsilon \over l^2}>{1 \over 18}$ is the inverse 
of the Hawking-Page phase transition, where the black hole is 
stable at high temperature but unstable at low 
temperature. The above condition
Eq.(\ref{phph}) in terms of AdS/CFT set-up  gives 
the bound for rank $N$  of gauge group for dual CFT.
For example, taking  group $Sp(N)$ of corresponding ${\cal N}=2$ 
dual superconformal theory (see description
of the model at the end of next section) one arrives to 
critical value $N=288$. 

Hence, the situation is the following. For large $N$ as we observed 
the free energy is always negative and AdS BH is stable. Then, only 
one phase (deconfinement) occurs for dual CFT living on such 
conformal boundary. On the same time, below the critical value on $N$
there appears possibility of the inverse of the Hawking-Page phase
 transition thanks 
to non-zero $c$ contribution. For dual theory it means
that confinement-deconfinement transition occurs at some temperature. 
New (confining) phase may appear but only below the critical $N$! This is 
purely higher curvature term effect. 
Note the conformal boundary of above 5d AdS is hyperbolic space.

\section{Phase transitions for  AdS black holes with 
Ricci-flat horizons\label{sec4}}

In the last section, in order to regularize the action when the 
black hole solution is substituted we subtract the action of 
the AdS vacuum. In case of $k=0$, however, there is an argument 
that it is easier to subtract the action of AdS soliton \cite{HM} 
instead of the vacuum AdS \cite{SSW}. (Physical results on phase transitions 
are not changing). In this case, 
besides the temperature, the area of the horizon becomes an independent 
parameter on which the thermodynamical quantities depend. 

We now start from the construction of AdS soliton in 
$R^2$-gravity. When $k=0$, the AdS BH  
(\ref{dAdS}) looks like
\bea
\label{k0bh}
ds_{\rm BH}^2 &=& - \e^{2\rho_{\rm BH}(r)} dt_{\rm BH}^2 
+ \e^{-2\rho_{\rm BH}(r)}dr^2
 + r^2\left(d\phi_{\rm BH}^2 + \sum_{i=1,2}
\left(dx^i\right)^2\right)\ , \nn
\e^{2\rho_{\rm BH}(r)}&=&{1 \over r^2}\left\{-\mu_{\rm BH} 
+ {r^4 \over l^2} + {2\mu_{\rm BH}^2\epsilon \over r^4}
\right\}\ ,\quad \epsilon \equiv c\kappa^2\ .
\eea
In (\ref{k0bh}), we choose a torus for the $k=0$ Einstein 
manifold for simplicity. The coordinates of the torus are 
$\phi_{\rm BH}$ and $\left\{x^1,x^2\right\}$. One assumes 
$\phi_{\rm BH}$ has a period of $\eta_{\rm BH}$:
$\phi_{\rm BH}\sim \phi_{\rm BH} + \eta_{\rm BH}$.
The black hole horizon $r_{\rm BH}$ and the Hawking temperature 
$T_{\rm BH}$ are given by
\be
\label{rBH}
r_{\rm BH}= l^{1 \over 2}\mu_{\rm BH}^{1 \over 4} 
 - {1 \over 2}\epsilon l^{-{3 \over 2}}
\mu_{\rm BH}^{1 \over 4}\ ,\quad
T_{\rm BH}= {1 \over 4\pi}\left\{{4r_{\rm BH} \over l^2} 
- {8\epsilon r_{\rm BH} \over l^4}\right\}\ .
\ee
The Hawking temperature gives the periodicity of the time 
coordinate $t_{\rm BH}$ when we analytically continue the time 
coordinate:
$it_{\rm BH}\sim it_{\rm BH} + {1 \over T_{\rm BH}}$.

The AdS soliton solution can be obtained by exchanging the 
signature of $t_{\rm BH}$ and $\phi_{\rm BH}$ as 
$t_{\rm BH}\rightarrow i\phi_s$ and 
$\phi_{\rm BH}\rightarrow it_s$.
Then the metric of the AdS soliton is given by
\bea
\label{k0sltn}
ds_s^2 &=& - r^2 dt_s + \e^{-2\rho_s(r)}dr^2
 + \e^{2\rho_s(r)} d\phi_s^2 + r^2\sum_{i=1,2}
\left(dx^i\right)^2\ , \nn
\e^{2\rho_s(r)}&=&{1 \over r^2}\left\{-\mu_s 
+ {r^4 \over l^2} + {2\mu_s^2\epsilon \over r^4}
\right\}\ .
\eea
In the solution the radial coordinate is restricted to be:
\be
\label{rs}
r\geq r_s\ ,\quad r_s= l^{1 \over 2}\mu_s^{1 \over 4} 
 - {1 \over 2}\epsilon l^{-{3 \over 2}}\mu_s^{1 \over 4}\ .
\ee
The regularity at $r=r_s$ determines the periodicity of 
$\phi_s$:
\be
\label{Ts}
\phi_s\sim \phi_s + {1 \over T_s}\ ,\quad 
T_s= {1 \over 4\pi}\left\{{4r_s \over l^2} 
- {8\epsilon r_s \over l^4}\right\}\ .
\ee
One also assumes that the time coordinate $t_s$ has a periodicity
$\eta_s$ when analytically continued:
$it_s\sim it_s + \eta_s$. 

The free energy of the AdS black hole may be calculated 
from the action as above. As in (\ref{regS}), 
we regularize the action $S$ (\ref{sr}), where the 
AdS black hole solution (\ref{k0bh}) is substituted, by cutting 
off the integral at a large radius $r_{\rm max}$ and subtracting the 
action where AdS soliton (\ref{k0sltn}) is substituted. 
Then 
\be
\label{regS2}
S_{\rm reg}= -  
\left[{\eta_{\rm BH} V_2 \over T_{\rm BH}}
\int_{r_{\rm BH}}^{r_{\rm max}} dr r^3
 - {\eta_s V_2 \over T_s}\int_{r_s}^{r_{\rm max}} dr r^3 
\right]\left({8 \over \kappa^2 l^2} 
 - {32 c \over l^4} - {48 c\mu_{\rm BH}^2 \over r^8}\right)\ .
\ee
Here $V_2$ expresses the volume corresponding to the 
coordinates $\left\{x^1,x^2\right\}$.  
We now impose the following matching conditions at $r=r_{\rm max}$, 
which guarantee 
that the length of $t_{\rm BH}$ ($\phi_{\rm BH}$) direction on the 
cutted boundary is identical with that of $t_s$ ($\phi_s$) 
direction: 
\be
\label{mcndtn}
{\e^{\rho_{\rm BH}(r_{\rm max})} \over T_{\rm BH}}
= r_{\rm max}\eta_s\ ,\quad 
r_{\rm max} \eta_{\rm BH} = {\e^{\rho_s(r_{\rm max})} 
\over T_s}\ .
\ee
The conditions (\ref{mcndtn}) determine the parameters $\mu_s$ 
($r_s$ or $T_s$) and $\eta_s$ in terms of the parameters 
$\mu_{\rm BH}$ ($r_{\rm BH}$ or $T_{\rm BH}$) and 
$\eta_{\rm BH}$. In the limit of $r_{\rm max}\rightarrow 
\infty$, one gets
\be
\label{mcndtn2}
\eta_s\rightarrow {1 \over lT_{\rm BH}}\ ,\quad 
T_s\rightarrow {1 \over l\eta_{\rm BH}}\ .
\ee
In the limit, we obtain the following expression of 
$S_{\rm reg}$: 
\be
\label{Sreg3}
S_{\rm reg}=-{\eta_{\rm BH} V_2 \over T_{\rm BH}}
\left[\left({1 \over \kappa^2} - {4c \over l^2} \right)
\left\{\mu_{\rm BH} - \mu_s 
 - 2 \left(r_{\rm BH}^4 - r_s^4\right) \right\}
 - 12c \left({\mu_{\rm BH}^2 \over r_{\rm BH}^4} 
 - {\mu_s^2 \over r_s^4}\right)\right]\ .
\ee
Using Eqs.(\ref{rBH}), (\ref{rs}), and (\ref{Ts}), one gets
\be
\label{rmu}
r_i=l^2\left(\pi T_i\right)\left( 1 + {2 \epsilon \over l^2}
\right)\ ,\quad 
\mu_i=l^6\left(\pi T_i\right)^4\left(1 + {10 \epsilon \over l^2} 
\right)\ ,\quad (i={\rm BH,\ s})\ .
\ee
By using (\ref{rmu})  and (\ref{mcndtn2}), we can rewrite the 
expression in (\ref{Sreg3}) and obtain the free energy $F$ in 
the following form:
\bea
\label{F}
F&=&-{\eta_{\rm BH} V_2 l^6 \over \kappa^2}\left(1 
+ {14\epsilon \over l^2}\right)\left\{\left(\pi T_{\rm BH}\right)^4 
 - \left(\pi T_s\right)^4 \right\} \nn
&=&-{\eta_{\rm BH} V_2 l^6 \over \kappa^2}\left(1 
+ {14\epsilon \over l^2}\right)\left\{\left(\pi T_{\rm BH}\right)^4 
 - \left({\pi \over l\eta_{\rm BH}}\right)^4 \right\} \ .
\eea
When $c=0$, the above expression reproduces the result 
in \cite{SSW}\footnote{
When $c=0$, the expression corresponding to Eq.(11) 
in \cite{SSW} 
\[
\beta_b F = I ={{\rm vol}({\cal F}) \over 16\pi G l} 
\beta_b \beta_s \left[{\bf k}^{n-1}_s 
 - {\bf k}^{n-1}_b\right]
\]
can be reproduced from (\ref{F}) by replacing
$V_2={\rm vol}({\cal F})$, $\kappa^2=16\pi G$, 
$\eta_{\rm BH}={\beta_s \over l}$, 
$l^6\left(\pi T_{\rm BH}\right)^4=\mu_{\rm BH}={\bf k}^{n-1}_b$, 
$l^6\left(\pi T_s\right)^4=\mu_s=l^6 
\left({\pi \over l\eta_{\rm BH}}\right)^4= {\bf k}^{n-1}_s$.}.
Eq.(\ref{F}) tells that there is a phase transion, as in $c=0$ 
case \cite{SSW}, at
\be
\label{Tc}
T_{\rm BH}={1 \over l\eta_{\rm BH}}\ .
\ee
If $1 + {14\epsilon \over l^2}>0$, the situation is not so 
changed from $c=0$ case and we find that 
when $T_{\rm BH}>{1 \over \eta_{\rm BH}}$, the black 
hole is stable but when $T_{\rm BH}<{1 \over \eta_{\rm BH}}$, 
the black hole becomes unstable and the AdS soliton  is 
preferred. Eq.(\ref{F}), however, suggests that there appears a 
critical point at 
\be
\label{epc}
1 = -{14\epsilon \over l^2}\ .
\ee
If $1 + {14\epsilon \over l^2}<0$, the situation is changed 
and we find that 
when $T_{\rm BH}>{1 \over \eta_{\rm BH}}$, the black 
hole is unstable and the AdS soliton  is 
preferred but when $T_{\rm BH}<{1 \over \eta_{\rm BH}}$, 
the black hole becomes stable. 
As we treat the correction from $\epsilon$ perturbatively, 
it is not  clear if the above critical point really occurs.
In order to prove this fact, 
 we consider the situation 
that $c=0$ but $a$ and/or $b$ do not vanish in the action 
(\ref{vi}). When $c=0$ but $a$, $b\neq 0$, the 
Schwarzschild-AdS solution and the AdS soliton solution 
are exact solutions, whose metric are given by replacing 
$\e^{2\rho_{\rm BH}(r)}$ in (\ref{k0bh}) and 
$\e^{2\rho_s(r)}$ in (\ref{k0sltn}) by
\be
\label{k0bh2}
\e^{2\rho_{\rm BH}(r)}\rightarrow {1 \over r^2}\left\{-\mu_{\rm BH} 
+ {r^4 \over l^2}\right\} \ , \quad
\e^{2\rho_s(r)}={1 \over r^2}\left\{-\mu_s 
+ {r^4 \over l^2} \right\}\ .
\ee
Then by the procedure similar to the case of $c\neq 0$, one 
finds the following expression for the free energy 
(analog of (\ref{F})):
\be
\label{F3}
F=- {\eta_{\rm BH} V_2 l^6 \over \kappa^2}
\left( 1 - {40 a\kappa^2 \over l^2}
 -{8 b\kappa^2 \over l^2} \right)
\left\{\left(\pi T_{\rm BH}\right)^4 
 - \left({\pi \over l\eta_{\rm BH}}\right)^4 \right\} \ .
\ee
We should note that the above expression (\ref{F3}) is valid for 
the arbitrary values of $a$ and $b$ although the expression 
(\ref{F}) is valid for small $\epsilon$ only. Eq.(\ref{F3}) tells 
again that there is a critical point (line), even for spherical AdS BH, when
\be
\label{abcrtcl}
 1 - {40 a\kappa^2 \over l^2}
 -{8 b\kappa^2 \over l^2}=0\ ,
\ee
which is an analog of (\ref{epc}). 
The above results demonstrate that there should appear a  
critical point in $R^2$-gravities. 

Let us consider the explicit example in
 the framework of AdS/CFT correspondence. 
The ${\cal N}=2$ theory with the gauge group $Sp(N)$ arises as 
the low-energy theory on the world volume on $N$ D3-branes 
sitting inside 8 D7-branes at an O7-plane \cite{Sen}. 
The string theory dual to this theory has been conjectured 
to be type IIB string theory on ${\rm AdS}_5\times {\rm X}^5$ where 
${\rm X}_5={\rm S}^5/Z_2$ \cite{FS}, whose low energy effective 
action is given by  
\be
\label{bng3} 
S=\int_{{\rm AdS}_5} d^5x \sqrt{G}\left\{{N^2 \over 4\pi^2}
\left(R-2\Lambda\right) 
+ {6N \over 24\cdot 16\pi^2}R_{\mu\nu\rho\sigma}
R^{\mu\nu\rho\sigma}\right\}\ .
\ee
Then $R^2$-term appears as $1/N$ correction. Identifying 
${\epsilon \over l^2}= {16 \over N}$, we find from 
(\ref{epc}) that there appears a critical point at 
\be
\label{epc2}
N = 160\ ,
\ee
where $N$ seems to be large enough.

Hence, the phase structure of such SCFT looks as following.
For large $N$ (above of critical value) there are two phases of
 AdS thermodynamics. Stable AdS BH phase corresponds to deconfinement 
of dual SCFT, at some critical temperature there occurs phase transition. 
The low temperature phase should 
correspond to the confining phase in the gauge theory from 
the viewpoint of the AdS/CFT correspondence.
When one considers the same theory but with $N$ less than critical value,
then the situation is reversed. What before was deconfinement becomes 
confinement and vice-versa. It is remarkable the phase transition occurs 
formally at the same critical temperature as above.
  Hence, even for 
the low temperature depending on $N$, there may occur and confinement and 
deconfinement. 
 
Hence, the role of next-to-leading correction in large $N$-
expansion (higher derivative term) is to clarify the structure of confining-
deconfining phases (and their reverse depending on $N$) of dual SCFT. 
Forgetting these corrections 
would lead to wrong conclusion about the phase transition at dual SCFT. 
This consideration suggests that taking account of further corrections 
(say $R^3$, $R^4$, etc) would make the phase structure even more complicated.
It could be that phase transitions with $N$ as order parameter may be 
observed in such framework. 

Indeed as toy example let us consider the following HD action 
including 
 say, $R^4$-term. 
For simplicity, we assume the Lagrangian density is given by the 
arbitrary scalar function $f(\hat G_{\mu\nu}, \hat R_{\mu\nu})$ 
of the metric $\hat G_{\mu\nu}$ and the Ricci tenosr 
$\hat R_{\mu\nu}$: 
\be
\label{Ldens}
S=\int d^{d+1} x \sqrt{-\hat G} f(\hat G_{\mu\nu},\hat R_{\mu\nu})\ . 
\ee 
The equation of motion derived from (\ref{Ldens}) has the 
following form:
\bea
\label{fEq}
0&=& {f \over 2}\hat G^{\mu\nu} 
+ {\partial f \over \partial \hat G_{\mu\nu}} 
+ {1 \over 2}\left(D_\rho D^\mu {\partial f \over 
\partial \hat R_{\rho\nu}} + D_\rho D^\nu {\partial f \over 
\partial \hat R_{\rho\mu}}\right) \nn
&& - {1 \over 2}D_\rho D^\rho {\partial f \over 
\partial \hat R_{\mu\nu}} - {1 \over 2}g^{\mu\nu}D_\rho D_\sigma 
{\partial f \over \partial \hat R_{\rho\sigma}} \ .
\eea
If we assume the metric is given by the Schwarzschild-de Sitter 
 (\ref{SAdS}), one finds that 
${\partial f \over \partial \hat G_{\mu\nu}}$ and 
${\partial f \over \partial \hat R_{\rho\sigma}}$ are proportional 
to $\hat G^{\mu\nu}$ with constant coefficients since the 
Ricci tensor $\hat R_{\mu\nu}$ is proportional to $\hat G_{\mu\nu}$ : 
$\hat R_{\mu\nu}=-{d \over l^2}\hat G_{\mu\nu}$. Since 
${\partial f \over \partial \hat G_{\mu\nu}}$ and 
${\partial f \over \partial \hat R_{\rho\sigma}}$ are covariantly 
constant, Eq.(\ref{fEq}) becomes simple: 
$0= {f \over 2}\hat G^{\mu\nu} 
+ {\partial f \over \partial \hat G_{\mu\nu}}$. 
By multiplying it to $G^{\mu\nu}$, one gets 
\be
\label{fEq3}
0= {d+1 \over 2}f 
+ G^{\mu\nu}{\partial f \over \partial \hat G_{\mu\nu}} \ .
\ee
Eq.(\ref{fEq3}) determines the length parameter $l^2$ in 
(\ref{SAdS}). Then the metric in the form of 
(\ref{SAdS}) is a solution of Eq.(\ref{fEq}). 
Substituting the metric (\ref{SAdS}) into the 
function $f(g_{\mu\nu},R_{\mu\nu})$, $f(g_{\mu\nu},R_{\mu\nu})$ 
becomes a constant $f(g_{\mu\nu},R_{\mu\nu})=f_0$
and does not depend on the mass parameter 
$\mu$. Using the method similar to that in Section \ref{sec2}, 
we find the following expression of the free energy for 
$d=4$ and $k\neq 0$ case:
\be
\label{Ff1}
F=-{V_3 \over 8}r_{H}^{2} \left( {r_{H}^{2} \over l^{2}}
 - {k \over 2} \right) f_0\ .
\ee
while for $d=4$ and $k=0$ case:
\be
\label{Ff2}
F=-{V_2\eta_{\rm BH} \over 8}r_{H}^{2} f_0 
\left\{\left(\pi T_{\rm BH}\right)^4 
 - \left(\pi T_s\right)^4 \right\}\ .
\ee
Eqs.(\ref{Ff1}) and (\ref{Ff2}) tell that there is a critical 
point (even for spherical AdS BH) when $f_0=0$. For $k=0$ case, as an 
example, when $f_0>0$ ($f_0<0$), the black hole is stable (unstable) if 
$T_{\rm BH}>{1 \over \eta_{\rm BH}}$ but global AdS 
becomes unstable (stable). 

To demonstrate the role of $R^4$-term let  us choose for $d=4$:
\be
\label{fexmpl}
f(g_{\mu\nu},R_{\mu\nu})
=a_1 R^2 + a_2 R^4 + {1 \over \kappa^2} R - \Lambda \ .
\ee
Then Eq.(\ref{fEq3}) has the following form:
\bea
\label{exmpl1}
0&=&{5 \over 2}\left(a_1 R^2 + a_2 R^4 + {1 \over \kappa^2} R
 - \Lambda \right) 
 - \left( 2 a_1 R^2 + 4 a_2 R^4 + {1 \over \kappa^2} 
R \right) \nn
&=& {200 a_1 \over l^4} - {240000 a_2 \over l^8} 
 -{80 \over \kappa^2 l^2} - {5 \over 2}\Lambda\ .
\eea
 and $f_0$ looks like
\be
\label{exmpl2}
f_0={400 a_1 \over l^4} + {160000 a_2 \over l^8} 
 - {40 \over \kappa^2 l^2} - \Lambda
={320 a_1 \over l^4} + {256000 a_2 \over l^8}
 - {8 \over \kappa^2 l^2}\ .
\ee
Here we have deleted $\Lambda$ by using (\ref{exmpl1}). 
Then there is a critical point when 
\be
\label{exmplcr}
0={40 a_1 \over l^4} + {32000 a_2 \over l^8}
 - {1 \over \kappa^2 l^2}\ .
\ee
The above result indicates to further modification of phase transition 
structure 
with account of higher powers of curvatures. In fact, in AdS/CFT set-up this 
suggests 
that above confinement-deconfinement phase transitions are deeply 
non-perturbative effect and (some?) non-perturbative technique 
should be used even when it is studied using SG dual description.

As a final remark let us note that above study may be
generalized to charged AdS BH thermodynamics where more complicated phase 
diagrams (see, for example,\cite{cg}) appear due to presence of extra 
parameter (charge).
Note added:  Next day this work was in hep-th, there appeared in hep-th
 ref.\cite{cai} where
similar question has been studied  for AdS BHs in
  Gauss-Bonnett theory.
It has been demonstrated there also that phase transitions depend on
Gauss-Bonnett parameter (in our case, this is combination of HD terms
 coefficients).

\section*{Acknowledgment}

We are grateful to M. Cveti\v c and I. Kogan for helpful discussions. 
The work by SDO has been supported in part by CONACyT (CP, Ref.990356).
The authors are indebted to I. Neupane since he pointed out 
the mistakes in the previous version.

\end{document}